\documentclass[%
reprint,onecolumn,
amsmath,amssymb,
aps,
prd,
]{revtex4-2}

\usepackage{amsmath}
\usepackage{comment}
\usepackage{appendix}
\usepackage{graphicx}

\begin{document}
\title{Resonant Axion--Photon Conversion in the Early Inspiral of Neutron Star Binaries}

\author{D. Su\'arez-Fontanella}
\email{duvier@usal.es}
\affiliation{
	Department of Fundamental Physics and IUFFyM,
	University of Salamanca,
	Plaza de la Merced SN,
	E--37008 Salamanca, Spain
}

\author{M. \'Angeles P\'erez-Garc\'ia}
\email{mperezga@usal.es}
\affiliation{
	Department of Fundamental Physics and IUFFyM,
	University of Salamanca,
	Plaza de la Merced SN,
	E--37008 Salamanca, Spain
}

\author{C. Albertus}
\email{albertus@usal.es}
\affiliation{
	Department of Fundamental Physics and IUFFyM,
	University of Salamanca,
	Plaza de la Merced SN,
	E--37008 Salamanca, Spain
}

\begin{abstract}

We consider the early binary neutron star inspiral phase as a scenario to probe environmental axion--photon resonant conversion. For this we approximately model the merger site electromagnetic fields as the superposition of two rotating dipolar stellar magnetic fields at the thousand--km scale when both magnetospheres are not largely distorted. We capture the time-sliced near-zone magnetospheric geometry relevant for axion--photon mixing. Plasma effects are incorporated through an effective Goldreich--Julian charge density, used to determine the effective plasma frequency and the location of resonant conversion surfaces. Our results show that axion--photon resonant conversion in binary magnetospheres mostly occurs on extended peanut-shaped surfaces whose global geometry evolves as the binary inspiral evolves. As a consequence, the total electromagnetic power emitted through axion--photon conversion exhibits a characteristic dependence on axion mass and a slow temporal modulation correlated with the gravitational wave frequency emission. This feature is potentially detectable for $m_a \in [50,170] \,\rm \mu eV$  and set $g_{a \gamma} \lesssim 10^{-11}\rm \,GeV^{-1}$ as it lies within the sensitivity limits of current or planned radio observation missions.
In light of our results we discuss the opportunity  of binary neutron star inspirals as time-dependent, multimessenger probes of axion physics, and motivate coordinated searches combining gravitational wave observations with radio and millimeter wavelength electromagnetic measurements.

\end{abstract}

\maketitle

\section{Introduction}

The nature of dark matter (DM) remains one of the central open problems in current fundamental physics and astrophysics. Among the candidates beyond the Standard Model, axions and, more generally, axion-like particles (ALPs) stand out due to their theoretical robustness  \citep{PecceiQuinn1977,Weinberg1978,Wilczek1978,Marsh2016} and their rich phenomenology across cosmological and astrophysical environments resulting in a vast zoo of them \citep{albertus2026wispediawispsencyclopedia}. In addition to laboratory experiments and direct searches \citep{IrastorzaRedondo2018,Adelberger2022}, indirect detection strategies based on astrophysical systems with extreme electromagnetic and plasma conditions provide a complementary avenue to probe axion interactions \citep{RaffeltStodolsky1988}. In particular, environments hosting strong magnetic fields and dilute plasmas offer natural settings for axion–photon conversion, potentially leading to observable electromagnetic signatures \citep{Sikivie1983}.

Compact binary coalescences and especially binary neutron star (BNS) systems, occupy a unique position in the multimessenger emission context. 
The global network of ground-based interferometers, including Advanced LIGO,
Advanced Virgo, and KAGRA, has enabled the routine detection of transient
gravitational wave (GW) signals from compact binary coalescences
\citep{Abbott2017,kagra}. Looking ahead, third-generation ground-based detectors such as the Einstein Telescope (ET) \citep{Punturo2010} and Cosmic Explorer(CE) \citep{reitze2019cosmicexploreruscontribution}, as
well as proposed space-based missions including DECIGO \citep{setoPhysRevLett.87.221103} and BBO, are expected to extend sensitivity to earlier inspiral phases and substantially increase the detection rate of BNS systems. These facilities will play a central role in future multi-messenger searches combining gravitational wave and electromagnetic observations.
At the same time, neutron stars (NS) possess some of the strongest magnetic fields known in the Universe, reaching surface intensities up to $B_s\sim 10^{15}\,\mathrm{G}$ in magnetars \citep{DuncanThompson1992,KaspiBeloborodov2017}. The coexistence of strong gravity, intense magnetization, and relativistic plasma dynamics make binary neutron stars sites natural multimessenger laboratories, where gravitational wave observations can be accompanied by electromagnetic and potentially dark-matter–induced signals \citep{MetzgerBerger2012, suarez2025gravitational, 2Suarez-Fontanella:2025tob}.

Focusing on the time evolution of the GW emission signal during the BNS merger event, it has been shown that during the inspiral phase, well before tidal disruption, the magnetospheres of the two neutron stars overlap and interact \citep{Lai2012,Piro2012} leading to a possible precursor signal \citep{Skiathas_2025}. Even in the absence of detailed magnetohydrodynamic modeling, the superposition of the individual stellar magnetic fields leads to a highly structured and time-dependent electromagnetic environment \citep{PhysRevD.88.043011}. Such configurations are expected to support extended regions where axion–photon conversion may occur resonantly, provided that the local plasma frequency matches the axion mass \citep{RaffeltStodolsky1988,Huang2018}. This mechanism does not rely on particle acceleration or non-thermal emission processes, but instead on coherent mixing in an external magnetic field, making it particularly sensitive to the global geometry of existing magnetic isocontours. 

In this work, we model the time-sliced conditions regarding the BNS site with explicit magnetospheric description focusing on the study of the resonant axion--photon conversion during the dynamical early inspiral phase. For this, the magnetic field is modeled at the thousand--km scale as the superposition of two rotating dipolar fields, capturing the near-zone magnetic geometry and safely recovering the leading-order vacuum solution \citep{Deutsch1955}  along with the magnetic structure assumed in standard magnetospheric plasma models \citep{GoldreichJulian1969,PhilippovSpitkovsky2018}. Thus assuming small magnetospheric distortion in the binary and while quasi-circular orbits are displayed,  we model the external plasma conditions by superposing local stellar Goldreich--Julian charge densities. This allows to approximately determine the resulting plasma frequency and the location of resonant conversion surfaces at fixed time.

Our calculation shows  that axion--photon conversion in this environment may efficiently occur on time evolving extended resonant peanut-shaped surfaces whose global geometry plays a central role in determining the emitted power as other multimessengers are produced as well. The dynamical evolution proceeds in a way that induces the total electromagnetic emission being periodically modulated, exhibiting a characteristic dependence on the axion mass that may be potentially detected with current or planned missions in the  GHz radio bands. Further, as the system evolves during the inspiral this modulation may be  naturally correlated with the gravitational wave signal. This identifies BNS inspirals as time-dependent, multimessenger probes of axion physics.

The organization of the paper is as follows. Section~\eqref{inspiral} introduces the electromagnetic model adopted for the binary system in the early inspiral phase, based on the superposition of two rotating dipolar stellar fields under smoothly distorted magnetospheric conditions.  In Sec.~\eqref{sec:axion_convertion} we analyze the plasma induced resonant axion--photon conversion in this magnetospheric configuration and derive the associated emitted power. In Sec.~\eqref{sec:observable} we study how the merger precursor conversion signal is modulated during the GW emission  and discuss its observability. Finally, we summarize our results and outline future directions in Sec.~\eqref{conclude}. We provide in the Appendix some technical derivations of mathematical expressions used. 
\section{Time-Sliced Binary Magnetospheres} \label{inspiral}

In this section we describe the model we assume for the effective electromagnetic description of the quasi-circular orbiting binary neutron star system in the early inspiral. This scenario is suitable for the analysis of a subsequently possible axion--photon resonant conversion in the magnetized site. The ultimate purpose of this time-sliced description is to provide a controlled and analytically tractable background magnetic field in the, otherwise, complex dynamics of the inspiral phase. We assume the separation between the two stars is large compared to their radii and the dynamics remains quasi-stationary from the perspective of the axion field. In this regime of weak coupling, the linearity of Maxwell’s equations implies that the total electromagnetic field can be approximated as the superposition of the individual stellar contributions.\\
We begin by recalling the electromagnetic field generated by a single rotating neutron star \citep{Deutsch1955} where each star is modeled as a perfectly conducting sphere of radius $a$, endowed with a magnetic field. Although a multipolar magnetic field is expected from stellar evolution and is therefore more realistic \citep{P_tri_2015}, a simplified description in terms of a rotating magnetic dipole inclined by an angle $\chi$ with respect to its spin axis remains widely used \citep{Gabler_2012, radice_binary}. We focus on the near-zone, defined by distances much smaller than the light-cylinder radius, $r \ll R_{\rm LC} = c/\omega$, where $\omega$ denotes the stellar angular velocity and $c$ is the speed of light. In this regime, the dipolar field solution admits explicit analytical expressions and captures the leading-order field structure relevant for magnetospheric dynamics and plasma processes.

In spherical coordinates $(r,\theta,\varphi)$, the magnetic field components at time $t$ generated by a single rotating neutron star with an oblique dipolar magnetic moment  were given by  \cite{Deutsch1955}, see Appendix for details. 

From this, we now consider a binary system composed of two similar neutron stars separated by a distance $d$. As long as the separation satisfies $d \ll R_{\rm LC}$, the system lies in the near-zone, quasi-static regime. Equivalently, this requires the characteristic light-crossing time across the binary, $d/c$, to be much shorter than both the spin and orbital timescales. 
As we will argue, in this limit, electromagnetic retardation and radiative effects are negligible, and Maxwell’s equations remain linear, with time dependence entering only parametrically through the rotational phases. The magnetic field generated by each star therefore retains the same functional form as in the isolated oblique-rotator case, and the total magnetic field of the binary can be constructed as a linear superposition of the individual stellar fields, once expressed in a common coordinate system. This approximation is expected to break down during the late inspiral,  when magnetospheric interactions and plasma effects induce non-linear couplings between the fields of the two NSs.\\
In the binary system, the total magnetic field in the meridional plane is
described as a superposition $\mathbf{B}_{\mathrm{binary}} = \mathbf{B} + \mathbf{B^{\star}}$ 
where the contribution $\mathbf{B^{\star}}$ corresponds to the displaced companion at a distance $d$ (see Appendix for further details).

At this stage, it is worth emphasizing that the above procedure concerns
exclusively the magnetic field configuration while no
assumptions are made regarding the associated electric field, which in realistic astrophysical conditions is not expected to vanish, although it is often neglected in practical calculations. Regardless of whether a vacuum description or a plasma-filled magnetosphere following the Goldreich--Julian (GJ) prescription \citep{GoldreichJulian1969} is adopted the near-zone magnetic field geometry will remain qualitatively similar and reasonably well captured by the rotating dipolar model considered here. Within the standard approach, the corresponding GJ electric field can then be obtained from the magnetic field as $\mathbf{E}_{\mathrm{GJ}} = - (\boldsymbol{\omega} \times \mathbf{r}) \times
\mathbf{B}$ up to corrections arising from plasma inertia and current closure effects.We assume stellar spin angular velocity governs corotation of the magnetospheric plasma. For completeness, the electric field components associated with the corresponding magnetic configuration in the vacuum dipole model are also provided in Appendix. \\
Finally, we note that in NS scenarios sufficiently dense axion clouds may develop on astrophysical timescales with saturation densities on the NS surface potentially exceeding $\mathcal{O}\left(10^{27}\right) \mathrm{GeV} \mathrm{cm}^{-3}$ for magnetars \citep{PhysRevX.14.041015, caputoPhysRevLett.133.161001}. In principle, the axion field may couple back to the electromagnetic sector, potentially leading to axion electrodynamics and modifying Maxwell’s equations  as backreaction effects can significantly alter the magnetospheric structure. However even for axion energy densities such as those mentioned, considering distance decline $\rho_a(r)\sim 1/r^4$ the axion--photon coupling remains perturbative $\epsilon \equiv g_{a \gamma} a_0 \sim g_{a \gamma} \frac{\sqrt{2 \rho_a}}{m_a} \sim 10^{-3}$ for experimentally allowed values of $g_{a\gamma}\lesssim 10^{-12}\,\rm GeV^{-1}$, implying that
axion-induced modifications to Maxwell's equations are negligible and that the underlying dipolar magnetospheric structure is preserved. Under this assumption, the electromagnetic fields derived above can be regarded as a fixed external background, and axion--photon conversion proceeds in the linear regime without modifying the underlying field structure.

\begin{figure}
\centering
\includegraphics[width=0.75\textwidth]{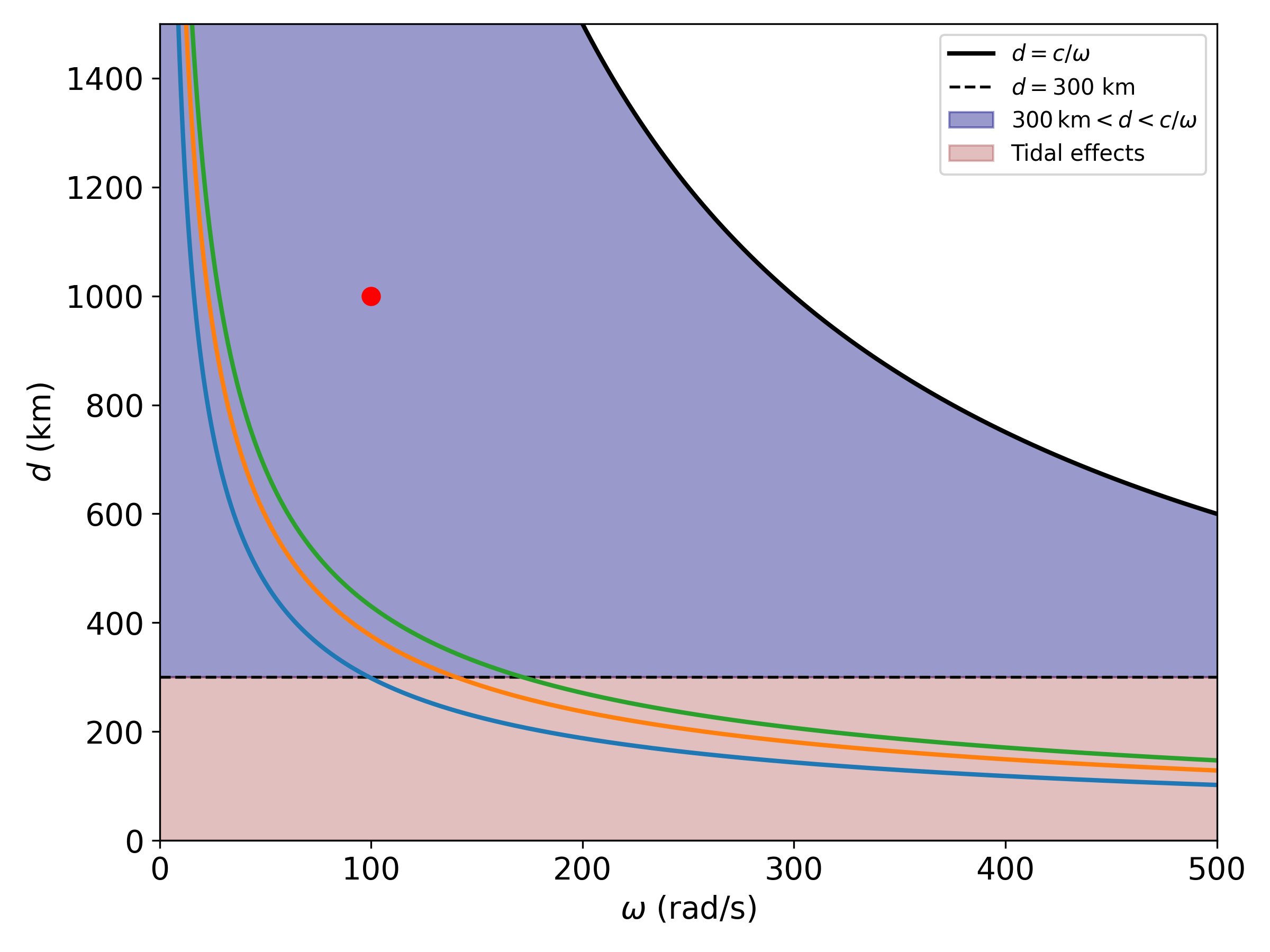}
\caption{%
Validity region of our model in the $d-\omega$ space. The purple area between solid and dashed black lines limits the boundary where our model's accuracy begins to degrade. In the lower pink band region tidal deformation effects start to manifest. Blue, orange and green curves denote orbital angular frequency $\Omega$ considering total BNS mass $M_\mathrm{tot}=$2,4, 6$M_\odot$ respectively, are superimposed on the same horizontal axes for the sake of comparison. The red dot shows our reference case at 1000 km.
}
\label{fig:validity-region}
\end{figure}

To demonstrate the validity of this idealized description, we present several
representative magnetic field configurations of the system.
\begin{figure}[t]
	\centering
	\includegraphics[width=0.307\textwidth]{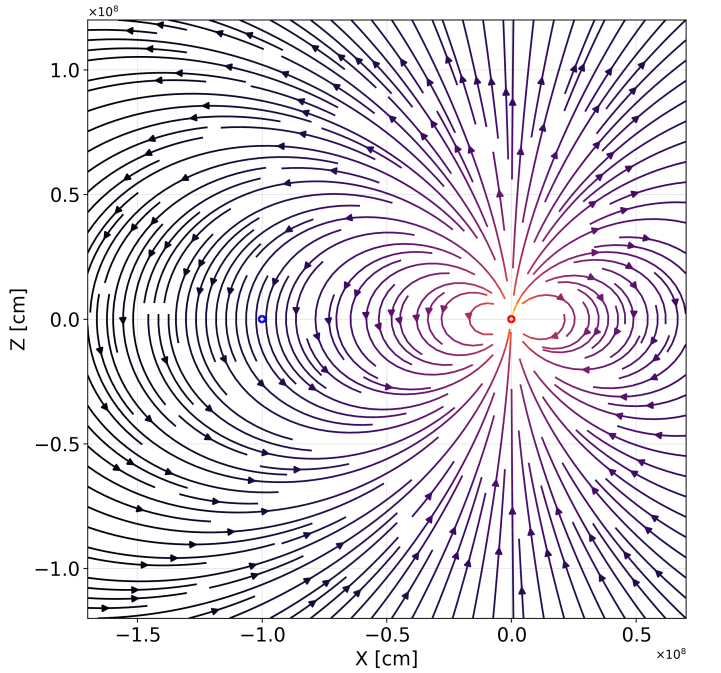}
	\hfill
	\includegraphics[width=0.273\textwidth]{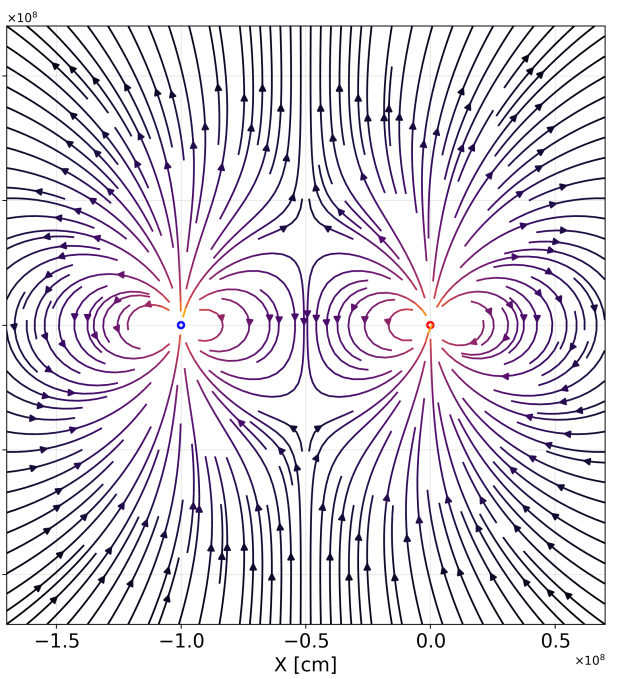}
	\hfill
	\includegraphics[width=0.32\textwidth]{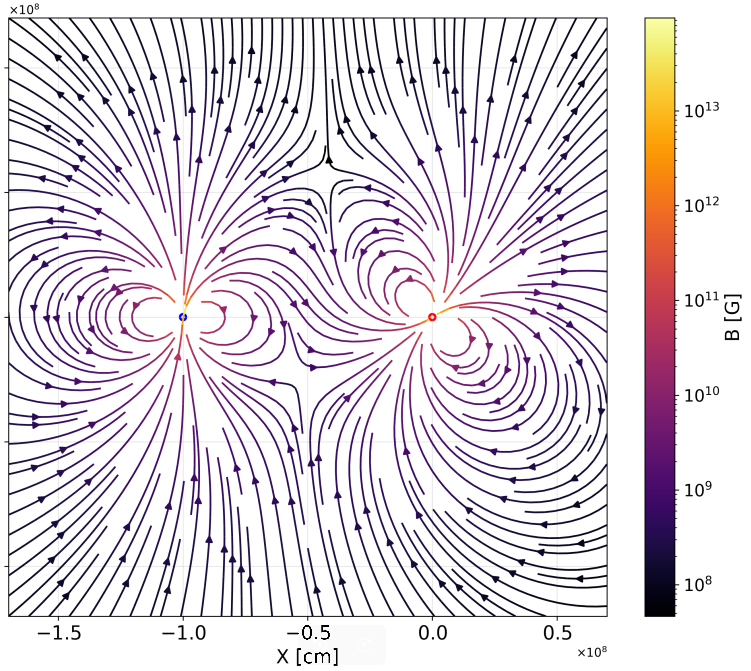}
	\caption{
		Configuration of magnetic field lines in the meridional plane of the binary neutron star system for time-sliced near-zone geometry. The panel shows the superposition of dipolar fields: a) isolated NS dipole b) superposition of both dipoles c) misaligned dipole moment for the right-hand star.
	}
	\label{fig:panel1}
\end{figure}
In Fig. \eqref{fig:panel1} we show the magnetic field configuration  for the BNS system in three different cases in the near-zone at time-sliced setting. On the left panel (a) it depicts a single NS case,  where  the magnetic field lines are perfectly symmetric and aligned with those of an expected dipolar field along the z-axis. In the central panel (b) we show the magnetic field configuration as both stars in the binary system contribute. This configuration illustrates how the magnetic fields overlap and influence each other due to their proximity. The interaction leads to a complex arrangement of field lines in the meridional plane even in absence of spin. Finally, in the right panel (c) we consider one of the NS magnetic field axis is tilted an angle $\frac{\pi}{4}$ relative to the z-axis. This inclination causes the magnetic field lines to fan out and deviate from the vertical orientation. The tilting introduces an asymmetry in the magnetic field distribution, resulting in a more complex pattern of field lines. This change illustrates how the inclination affects the magnetic interaction between the two stars, potentially leading to varied magnetic field strengths and orientations in the region surrounding the binary system.\\
We take a reference distance of $\sim1000 ~ \rm km $ between the stars that, as mentioned, remains consistent for our magnetic field treatment  in the near-zone $r \ll c/\omega$ for frequencies $\omega\in [0.1,1000]$ $\rm Hz$. 
On the other hand, tidal interactions become non-negligible once the binary separation decreases to a few hundred kilometers, typically $d \lesssim 300,\mathrm{km}$ for neutron star binaries. Below this scale, tidal deformations largely modify the quasi-circular orbital dynamics and the gravitational wave phase evolution in a way that must be accounted for using improved descriptions, see \citep{FlanaganHinderer2008}. This fact lead us  to set a region of model validity attending to the spin of each object that we illustrate in Fig. \eqref{fig:validity-region}

Although each NS in the binary possesses a non-negligible spin, with typical angular velocities of order $\sim10^{2}\,\mathrm{rad\,s^{-1}}$, the rotational contribution to the post-Newtonian (PN) dynamics of the system is small at the orbital separations considered here. For a representative neutron star with $m\simeq 1.4\,M_{\odot}$, $R\simeq 12\,\mathrm{km}$, moment of inertia $I\simeq kmR^2$ and $\omega\simeq 100\,\mathrm{rad\,s^{-1}}$, the corresponding dimensionless spin parameter is $\chi =k \frac{c R^2 \omega}{G m} \approx 8.1 \times 10^{-3}$. At a separation of $d = 1000\,\mathrm{km}$, the orbital angular frequency is
$\Omega = \sqrt{\frac{G M_{\mathrm{tot}}}{d^{3}}} \simeq 19.28\ \mathrm{rad\,s^{-1}},$
for a total mass $M_{\mathrm{tot}} = 2.8\,M_{\odot}$, yielding an orbital frequency of $f_{\mathrm{orb}} \simeq 3.07\,\mathrm{Hz}$ and a GW frequency of $f_{\mathrm{GW}}\simeq 6.14\,\mathrm{Hz}$. The orbital velocity at this distance is $v/c \simeq 0.0643$, so that the leading spin–orbit correction, which enters at 1.5PN order \citep{kidder1995coalescing, blanchet2014gravitational}, is suppressed by a factor $\chi (v/c)^{3} \approx 2.2\times 10^{-6}$. This value is roughly six orders of magnitude smaller than the Newtonian quadrupolar term, assuring that the influence of the stellar spins on both the GW frequency and the radiated power is entirely negligible in this regime.
On the other hand, from the axion perspective, the system should be quasi-static so the frequency associated to the axion field mass $m_a$ should be higher than the typical frequency considered in the binary system. For reference masses $m_a \lesssim 10^{-10}\,\mathrm{eV}$, the Compton frequency, $\omega_a=\frac{m_a c^2}{\hbar}$, $\nu_a=\frac{\omega_a}{2 \pi}\simeq 2.4\times10^{4}\,\mathrm{Hz}\left(\frac{m_a}{10^{-10}\,\mathrm{eV}}\right),$ is much larger than the characteristic orbital and spin frequencies of the system, $f_{\rm orb}\sim\mathcal{O}(1)\,\mathrm{Hz}$ and $f=\frac{\omega}{2\pi}\sim\mathcal{O}(10)\,\mathrm{Hz}$.
The electromagnetic background can therefore be treated as quasi-static during the conversion process. Spatial variations associated to the non relativistic axion field are sized by the de Broglie wavelength
$\lambda_{\rm dB}\simeq 2\pi/(m_a v_a)$ so that  for $m_a \sim 10^{-10} \mathrm{eV}$ yield $\frac{\lambda_a}{d} \gg 1 $ ensuring a locally smooth background.

\begin{figure}[ht]
	\centering
	\includegraphics[width=0.75\textwidth]{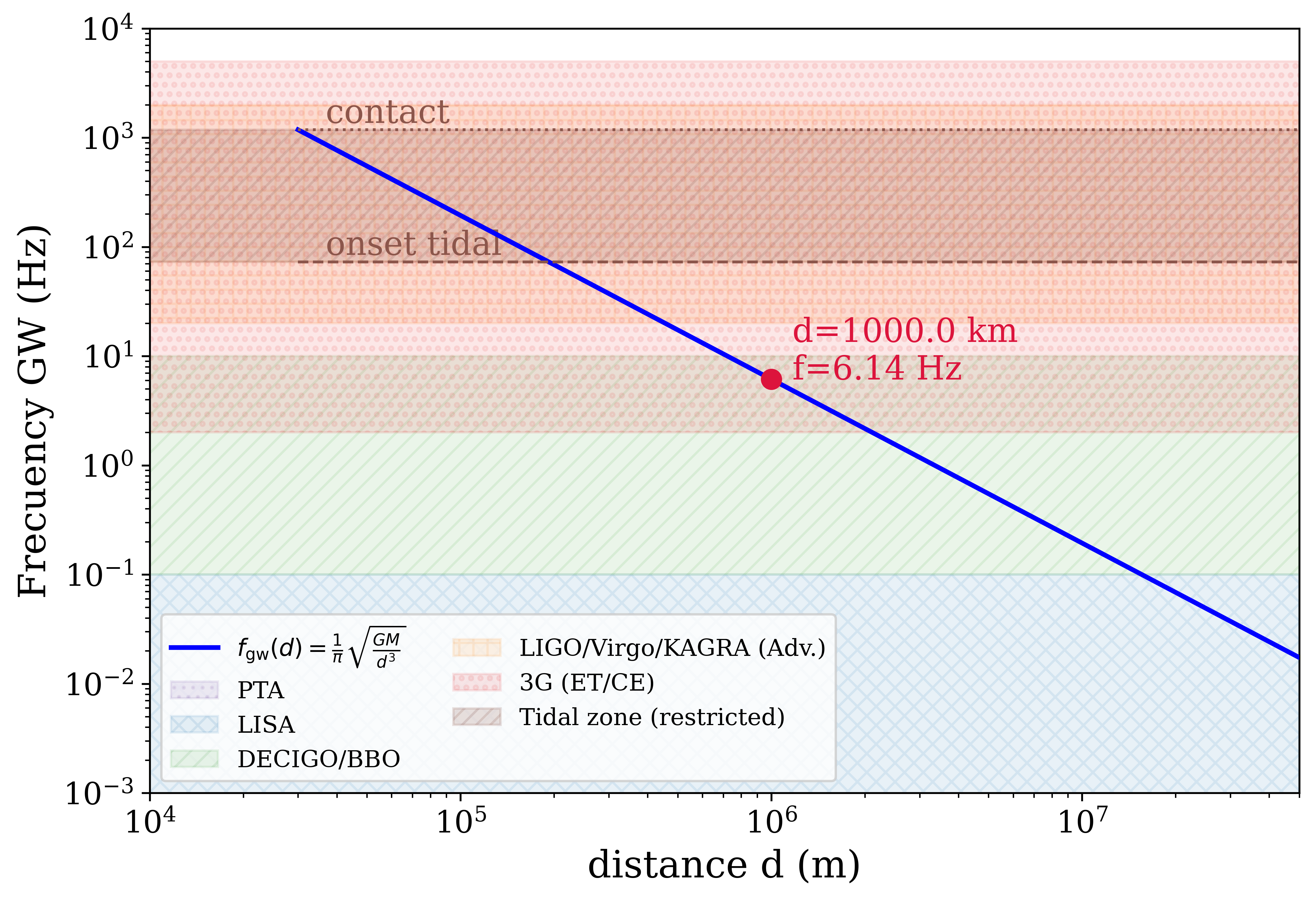}
	\caption{ 
		Gravitational wave frequency as a function of orbital separation for a
		quasi-circular BNS system. The blue curve shows the Keplerian
		relation, while the red point marks the reference case $d = 1000\,\mathrm{km}$, $f_{\mathrm{GW}} \simeq 6.14\,\mathrm{Hz}$ for two
		$1.4\,M_{\odot}$ NSs) belonging to the DECIGO/BBO and 3G (ET/CE) band. Shaded regions indicate the approximate
		sensitivity bands of current and future detectors, with the tidal regime
		excluded.}
	\label{fig:interferometers}
\end{figure}

Under these conditions, axion--photon conversion proceeds through localized resonant level crossings in the plasma filling the BNS site. The local plasma frequency will eventually match that of the axion field, $\omega_p(\mathbf{x})=\omega_a$. As we will explain in the following section, the conversion probability in this scenario is controlled by the local transverse magnetic field $\mathbf{B}_\perp$ and by the plasma frequency gradient normal to the resonant surface, according to the Landau--Zener formalism \citep{Landau1932,Zener1932}.

In order to connect the regime of validity of our model with the observational context, we already stated earlier in this work that the GW frequency related is  the orbital separation. Thus relation allows us to map  the frequency bands probed by current and future GW interferometers as shown in Fig.~\eqref{fig:interferometers}. It includes the current ground-based detector network formed by LIGO, Virgo, and KAGRA~\citep{Abbott2020}, as well as next-generation facilities such as the Einstein Telescope~\citep{Punturo2010} or Cosmic Explorer \citep{reitze2019cosmicexploreruscontribution}, the space-based mission LISA~\citep{AmaroSeoane2017}, and the proposed decihertz interferometer DECIGO~\citep{Kawamura2011}. The characteristic sensitivity curves of these instruments span several orders of magnitude in frequency, implying that the detectability of a given system is determined by whether its GW emission sweeps through the corresponding observational window.

\section{ Resonant Axion--Photon Conversion in the early Binary Magnetospheric configuration}
\label{sec:axion_convertion}

Building directly on the magnetic configuration constructed in the previous
section, we now address resonant axion--photon conversion in the magnetized
environment of the early inspiraling neutron star binary. In this setting, the plasma
properties entering the axion--photon dynamics are encoded in the
Goldreich--Julian (GJ) charge density \citep{GoldreichJulian1969}. In the
quasi-static regime considered here, we approximate the GJ density as the sum of
the individual stellar contributions,
\begin{equation}
n_{\mathrm{GJ}}(\mathbf{x}) \propto
\boldsymbol{\omega}_1 \cdot \mathbf{B}_1(\mathbf{x})
+
\boldsymbol{\omega}_2 \cdot \mathbf{B}_2(\mathbf{x}),
\end{equation}
where $\boldsymbol{\omega}_i$ denotes the spin angular frequency vector of NS 
$i=1,2$, and $\mathbf{B}_i$ its associated magnetic field. This additive structure
follows directly from the linearity of Maxwell's equations in the weak-field
regime and from the superposition principle adopted throughout our analysis.\\
It is worth mentionin at this point that in a binary, a global corotating frame does not exist, and the plasma cannot rigidly corotate with both stars simultaneously. In the present context, however, the GJ density should be interpreted as a local and instantaneous quantity associated with each star separately, defined in the near-zone where the magnetic field is dominated by a single dipole and varies slowly compared to the axion oscillation time scale. 
Within this interpretation, $n_{\rm GJ}(\mathbf{x})$ provides an effective estimate of the local charge density sourcing the plasma frequency, rather than a statement about the global dynamical state of the magnetosphere. \\
Since resonant axion--photon conversion depends only on the instantaneous value of $\omega_p(\mathbf{x})$ and not on whether the plasma is in exact corotation on large scales, treating the total charge density as explained,  captures the leading-order spatial structure relevant for locating the resonant surfaces. Deviations from exact corotation or magnetospheric interaction effects are therefore expected to affect primarily the detailed plasma dynamics, while leaving the geometry and inspiral-driven evolution of the resonant regions essentially unchanged.\\
In a dispersive plasma, electromagnetic waves acquire an effective mass set by the plasma frequency, so that the axion--photon system behaves as a
two-level system with a position-dependent mass splitting. Resonant axion--photon conversion occurs when the axion mass matches the effective photon mass induced by the plasma,
\begin{equation}\label{resonance_condition}
m_a^2 = \omega_{\mathrm{p}}^2(x) \;=\; \frac{4\pi \alpha}{m_e}\, n_e(x)\, ,
\end{equation}
where $n_e$ and $m_e$ are the electron number density and it mass,  respectively and $\alpha$ is the fine structure constant. In terms of an effective photon mass $m_\gamma^2(x) \simeq \omega_p^2(x)$. 
More generally, the effective photon mass can be written as
\begin{equation}
m_\gamma^2(x) = \omega_p^2(x) + \delta m_\gamma^2(B)\, ,
\end{equation}
where $\delta m_\gamma^2(B)$ accounts for magnetic-field–induced corrections \citep{remark2023}. For the magnetic field strengths at resonant sites considered
here, $B \lesssim 10^{13}\,\mathrm G$, this term remains perturbative and does not alter the location of the resonant conversion region. Typically in the plasma regions $\omega_p^2 \gg \delta m_\gamma^2(B)$.

\begin{figure}[t!]
	\centering
	\includegraphics[width=0.75\textwidth]{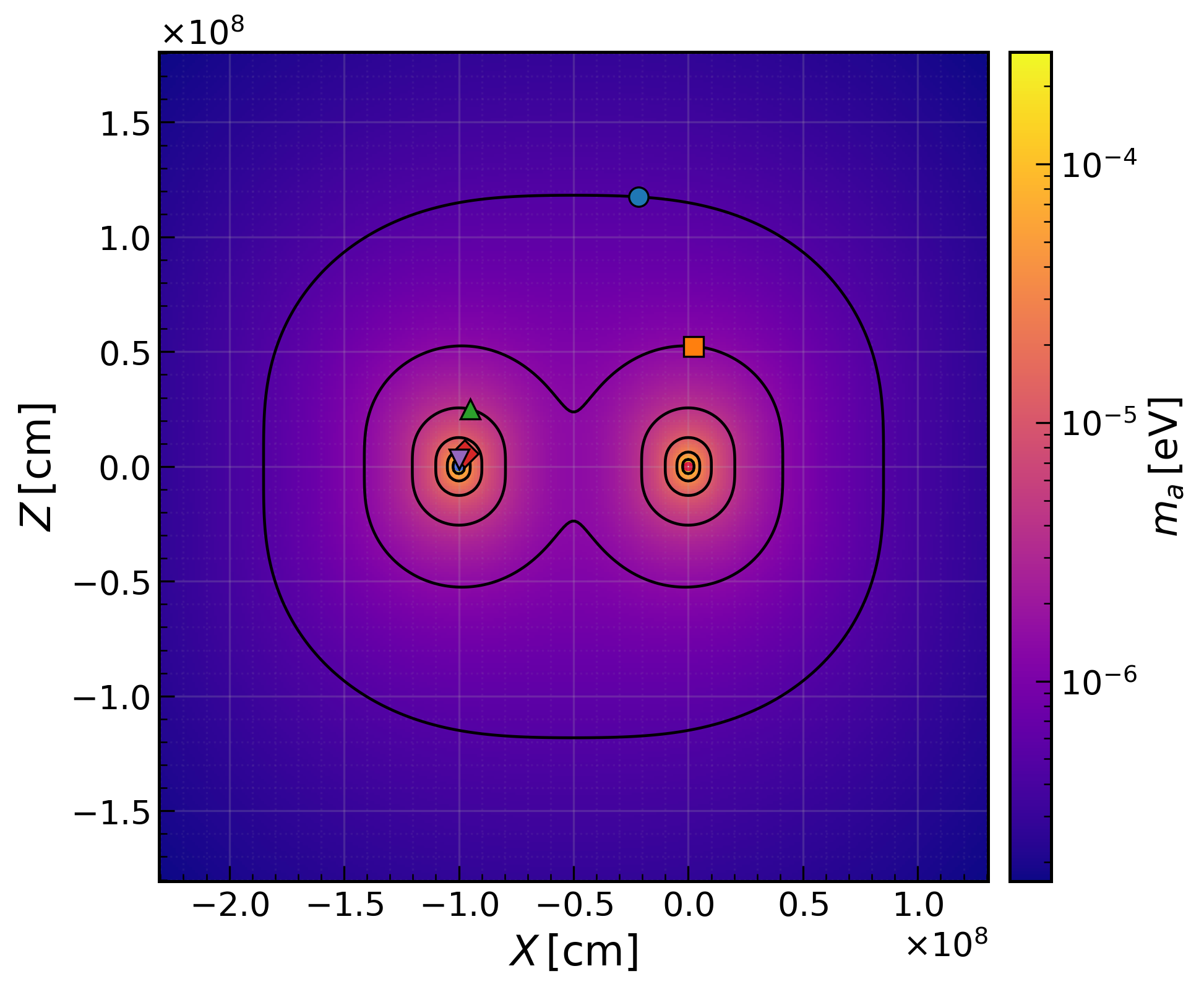}
	\caption{Resonant 2D isocontours for a binary system of two NSs with aligned spins and parallel magnetic fields, sharing the same spin frequency $\omega$ and surface magnetic field strength $B_S=10^{14}$ G. Each isocontour corresponds to a fixed axion mass at resonance. Arc length associated with each resonant contour for selected cases: blue circle $m_a=0.58 \, \rm \mu eV,\,L=7600 \rm \, km$, orange squared $m_a=1.4 \, \rm \mu eV, \,L=5400\, \rm km$, green triangle $m_a=5.5 \, \rm \mu eV, L=2300\, \rm km$), red squared $m_a=23 \, \rm \mu eV,\, L=929\, \rm km$, purple triangle $m_a=92 \, \rm \mu eV,\, L=365\, \rm km$.}
	\label{fig:resonance_curve}
\end{figure}

In the magnetospheric context considered here, we identify $n_e$ with the Goldreich--Julian density $n_{\mathrm{GJ}}$ up to a multiplicity factor, so that the resonant axion mass can be spatially scanned within this range. As a result, isocontours of constant $n_{\mathrm{GJ}}$ coincide exactly with isocontours of constant resonant axion mass in the meridional plane depicted in  Fig. \eqref{fig:resonance_curve}.
This fact has an important geometrical implication since for a given axion mass, the resonance condition is satisfied along closed boundary curves in space defined by constant  $n_{\mathrm{GJ}}(\mathbf{x})$ value. In the axisymmetric slices considered in our numerical analysis, these curves appear under a variety of forms, see circle, square or triangles in the interbinary region. Resonant conversion therefore takes place over the corresponding extended surfaces as shown in  Fig. \eqref{fig:resonance_curve} for a binary system of two NSs with aligned spins and parallel magnetic fields, sharing the same spin angular frequency $\omega$ and surface magnetic field strength $B_S=10^{14}$ G. Each depicted isocontour corresponds to a fixed axion mass at resonance and define an associated arc length. From inspection, resonant surface geometry depends sensitively on the axion mass and on the binary magnetic configuration.

The local probability for axion--photon conversion at resonance can be expressed in a Landau--Zener type form \citep{Carenza_2023},

\begin{equation}
P_{a\rightarrow\gamma}(\mathbf{x})
\simeq
1-\exp\!\left[
-\pi\,
\frac{g_{a\gamma}^2\,B_\perp^2(\mathbf{x})\,\omega_a}
{v_\perp(\mathbf{x})\,
\left|\partial_\ell \omega_p^2(\mathbf{x})\right|_{\mathbf{x}\in\mathcal{S}_{\rm res}}}
\right],
\qquad \mathbf{x}\in\mathcal{S}_{\rm res}(m_a),
\end{equation}

where $g_{a\gamma}$ is the axion--photon coupling, $\mathbf{B}_\perp$ denotes the component of the magnetic field transverse to the axion propagation direction. 
The derivative $\partial_\ell \omega_{\mathrm{p}}^2$ represents the spatial gradient of the plasma frequency squared evaluated along the direction normal to the resonant surface $\mathcal{S}_{\rm res}$. The velocity factor $v_\perp$ arises from converting this spatial gradient into the effective time rate at which the resonance is crossed along the axion trajectory, i.e. $d/dt = v_\perp \partial_\ell$. 

In this work, we approximate the axion trajectories as purely radial paths originating around the NS as this captures the leading-order contribution to the resonant conversion and does not imply relativistic axion velocities. Rather, it is compatible with our assumed scenario in which axions constitute a gravitationally bound and long-lived dark-matter cloud surrounding the NS, remaining stable over astrophysical timescales of millions of years\citep{PhysRevX.14.041015}. A fully consistent treatment would require integrating the conversion probability over the angular distribution of incoming trajectories, which we leave for future work.\\
The total electromagnetic power emitted through resonant axion–photon conversion is obtained by integrating the local conversion probability over the set of resonant surfaces $\mathcal{S}_{\rm res}(m_a)$ defined by Eq. \eqref{resonance_condition}, weighted by the incident axion flux. For a given axion mass, the resulting power can be written as

\begin{equation}
P_{\mathrm{tot,a\gamma}}\left(m_a\right)=\int_{\mathcal{C}\left(m_a\right)}(2 \pi \bar{r}) d \ell n_a(\mathbf{x}) v_{\perp}(\mathbf{x})\left(\hbar \omega_a\right) P_{a \rightarrow \gamma}(\mathbf{x})
\label{power3d}
\end{equation}

where the effective surface on the axisymmetric curve is $d A=2 \pi \bar{r} d \ell$ and $\bar{r}=r\sin\theta$ is the cylindrical radius.   In addition $n_a=\rho_a/m_a$. The  curve $\mathcal{C}(m_a)$ is the contour of the resonant
surface $\mathcal{S}_{\rm res}(m_a)$. If it has multiple disconnected components, the total power is obtained by summing their $k$ contributions,
$\int_{\mathcal{C}} \rightarrow \sum_k \int_{\mathcal{C}_k(m_a)}$.

The emitted power therefore arises from a nontrivial interplay among three key ingredients: the axion mass, which fixes both the axion energy and the spatial location of the resonance via Eq. \eqref{resonance_condition}; the local magnetic field, which determines the conversion efficiency; and the geometry of the resonant region, encoded in the total length of the resonant isocontours. While the magnetic field strength largely sets the overall normalization of the signal, geometric effects are essential in shaping its dependence on the axion mass.

For sufficiently large axion masses, the resonance condition selects small, compact isocontours localized deep within the inner magnetosphere of each NS (triangles and red square in Fig.~\eqref{fig:resonance_curve}). In this regime, resonant conversion occurs in regions of very strong magnetic field, leading to high local conversion probabilities. However, the resonant surfaces are geometrically limited and remain disconnected, each surrounding a single star. 

As the axion mass is decreased, the resonant isocontours lie distant to the central stellar site and their total length increases. In the intermediate-mass regime, this expansion is accompanied by a sharp geometric transition: previously disconnected, star-centered resonant contours such as that of green triangle, merge into a single, extended structure spanning the inter object region (blue circle or orange square in Fig.~\eqref{fig:resonance_curve}). 


\section{Inspiral-driven modulation of the Resonant  axion--photon conversion}
\label{sec:observable}

As shown in Sec.~\eqref{sec:axion_convertion}, the total power emitted through resonant axion--photon conversion is obtained by integrating the local Landau--Zener conversion probability over the set of resonant isocontours as inspiral evolves. 

\begin{figure}[t!]
\centering
\includegraphics[width=0.75\textwidth]{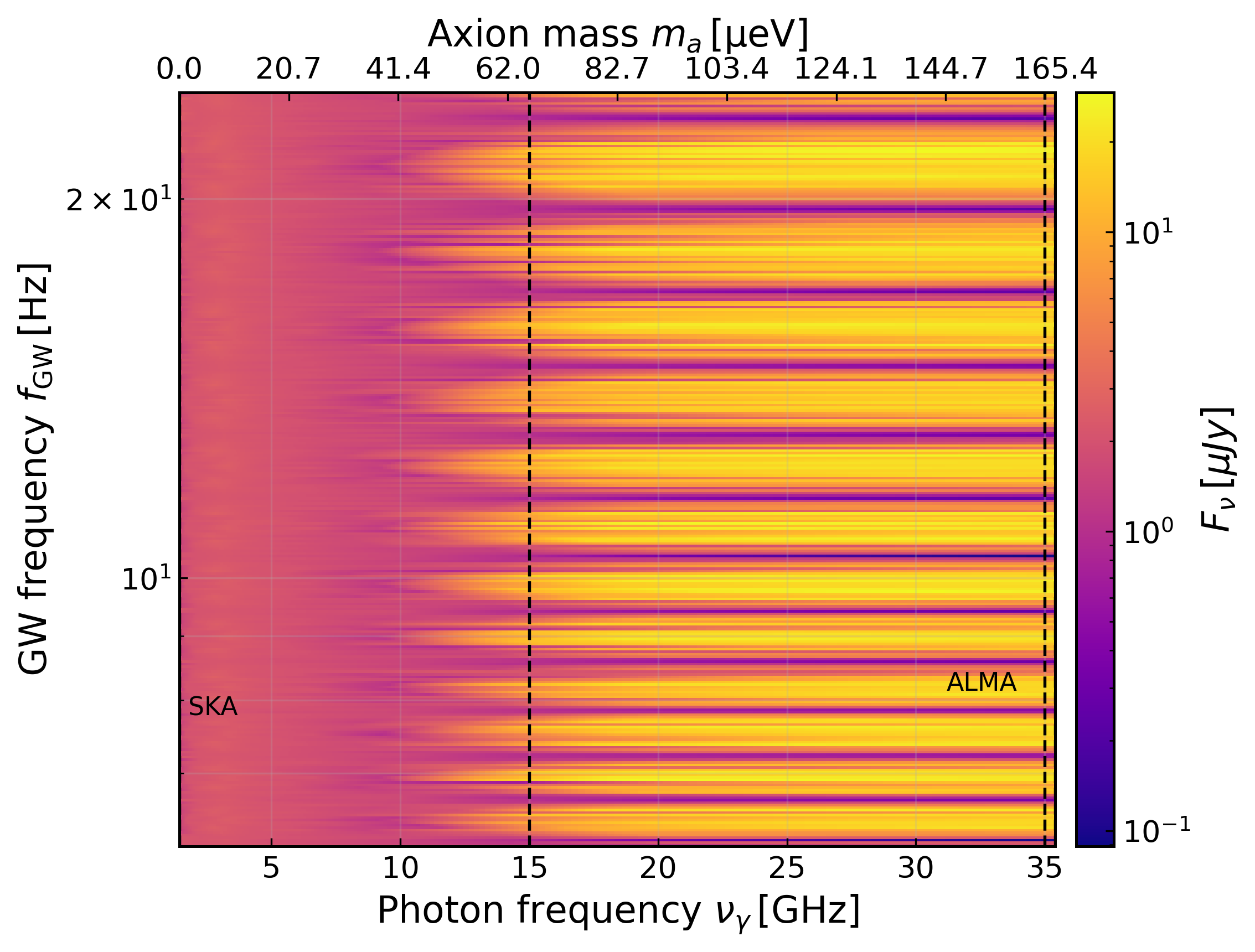}
\caption{  Predicted spectral flux density $F_{\nu,a\gamma}$ as a function of the photon frequency $\nu_\gamma$ and the GW frequency $f_{\mathrm{GW}}$ (from $d=1000\,\mathrm{km}$ to $400\,\mathrm{km}$) for a source distance of $10\,\mathrm{kpc}$. We set $g_{a \gamma}=10^{-12}\rm \,GeV^{-1}$ and  $\rho_a\sim 10^{23}\, \rm GeV cm^{-3}$ for this plot. Vertical dashed lines mark the
	representative observing bands of radio and millimeter facilities,  SKA (from 0.05 to 15 GHz)  and ALMA (from 35 to 50 GHz). VLA (not shown) covers from 0.01 to 50 GHz. The band-like 
	structures reflect the precursor signal in the inspiral-driven modulation of the resonant axion--photon conversion signal as successive resonant surfaces appear and
	disappear during the binary evolution.}
\label{fig:Fnu_map}
\end{figure}
The physical picture is as follows. As the two NSs approach each other in the quasi-circular orbits, the characteristic spatial extent of the magnetospheric regions, and hence of the resonant conversion surfaces, is set by the binary separation during the inspiral. The total emitted power given in Eq.\eqref{power3d} will act as a measure of the size of the resonant surfaces.

The quantity directly relevant for electromagnetic observations is the spectral flux density received at Earth. Assuming that the fraction of converted photons escape the magnetospheric regions and propagate freely to the observer, the observable flux density can be written as
\begin{align}
	F_{\nu,a\gamma} (\nu_\gamma,f_{\rm GW}) = \frac{P_{\mathrm{tot,a\gamma}}}{4\pi D^2\,\Delta\nu},
	\label{eq:Fnu_def}
\end{align}
where $D$ denotes the luminosity distance from observer to the source and $\Delta\nu$ is the intrinsic linewidth of the signal set primarily by the axion velocity
dispersion and by spatial variations of the plasma frequency across the
resonant conversion region.

Although the axion mass (and hence the emitted photon frequency
$\nu_\gamma\simeq m_a c^2/h$) is fixed, the observed spectral flux density varies
significantly during the inspiral. This modulation reflects the evolution of the
magnetospheric plasma and magnetic-field geometry as the binary separation
decreases. As the system evolves, the resonance condition
$\omega_p(\mathbf{x})=\omega_a$ is satisfied on different spatial surfaces whose
location, topology, and effective area change with time rather than from any intrinsic time-dependence of the axion frequency. We show the band-like
structures appearing in Fig.\eqref{fig:Fnu_map} for the predicted spectral flux density $F_\nu$ as a function of the photon frequency $\nu_\gamma$ from axion conversion and the GW frequency $f_{\mathrm{GW}}$ for each time instant. We scan from $d=1000\,\mathrm{km}$ to $400\,\mathrm{km}$ for a source distance of
$D=10\,\mathrm{kpc}$. Vertical dashed lines mark the representative observing bands of radio and millimeter facilities Square Kilometre Array (SKA) \citep{SKA_science} (from 0.05 to 15 GHz)  and the Atacama Large Millimeter/submillimeter Array (ALMA) \citep{ALMA_sensitivity,ALMA_Primer} (from 35 to 50 GHz). VLA \citep{VLA_sensitivity} covers from 0.01 to 50 GHz.  The band-like structures reflect the inspiral-driven modulation of the resonant axion--photon conversion signal as successive resonant surfaces appear and
We scan accordingly associated axion mass and photon coupling  values are currently not excluded \cite{cajohare_axionlimits}.

The emission produced by axion--photon conversion is spectrally narrow. Although each converted photon carries an energy $\omega_\gamma\simeq\omega_a=m_a$, the axion population possesses a finite velocity dispersion. For non-relativistic axions with typical velocities $v_a\sim10^{-3}$, this induces a Doppler broadening $\frac{\Delta\nu}{\nu_\gamma}\sim v_a ,$ so that $\Delta\nu\simeq\nu_\gamma v_a$. Equation~\eqref{eq:Fnu_def} therefore yields a well-defined prediction for the spectral flux density which can be directly compared with radio and millimeter observations. For reference, the commonly used radioastronomical unit is the Jansky, $1~\mathrm{Jy}=10^{-23}\,\mathrm{erg\,cm^{-2}\,s^{-1}\,Hz^{-1}}$.
Using the quasi-circular approximation for the early inspiral frequency, each magnetospheric configuration and its associated resonant geometry can be uniquely mapped to a given GW frequency. This one-to-one correspondence enables the construction of a two-dimensional observable $F_{\nu,a\gamma} (\nu_\gamma,f_{\rm GW})$. 
Fig.\eqref{fig:Fnu_map} exhibits a characteristic band-like structure as a function of the GW frequency. These features arise from the appearance and disappearance of extended resonant surfaces as the binary separation decreases during the early inspiral. While the overall normalization of the signal scales with the local axion density and the axion--photon coupling, the morphology of the modulation is controlled by the plasma frequency gradients and by the global geometry of the magnetic field. 
Importantly, the modulation of the electromagnetic signal with $f_{\rm GW}$ constitutes a robust and distinctive multimessenger signature of resonant axion--photon conversion in binary magnetospheres. Standard electromagnetic emission from neutron star binaries is typically broadband, including radio pulsar emission, thermal surface emission peaking in the X-ray band, and smooth non-thermal magnetospheric radiation. None of these mechanisms generate narrow spectral features in the radio or millimeter bands that are correlated with the GW  frequency. By contrast, the cool-axion-induced signal considered here appears as a narrow spectral feature whose frequency is fixed by the axion mass and whose amplitude evolves predictably with the inspiral, providing a clear observational discriminant against conventional electromagnetic backgrounds.

The predicted values of $F_{\nu,a\gamma}(\nu_\gamma,f_{\rm GW})$ can be directly
compared with the sensitivity ranges of current and upcoming radio and
millimeter observatories. In the GHz band, the VLA reaches rms sensitivities at
the level of tens of microJansky in deep integrations, while the SKA is expected
to achieve sub-microJansky sensitivities in targeted observations. At higher
frequencies, ALMA provides excellent sensitivity to narrow spectral features in
the $35$--$950~\mathrm{GHz}$ range, with performance depending on bandwidth and
integration time. Although the axion-induced fluxes are generally small, the
intrinsically narrow spectral nature of the signal and its correlation with a
known gravitational-wave trigger significantly enhance its observability
relative to blind electromagnetic surveys.

Detectability is controlled by the contrast between the axion-induced narrow
spectral feature and the underlying continuum emission from the merger
environment, as well as by the instrumental sensitivity per spectral channel. A
convenient measure of this contrast is the fractional contribution
\begin{equation}
	\delta \equiv \frac{F_{\nu,a\gamma}(\nu_\gamma,f_{\rm GW})}{F_\nu^{\rm bg}},
\end{equation}
evaluated within a frequency channel of width $\Delta\nu$ comparable to the
intrinsic linewidth of the signal. The background flux density $F_\nu^{\mathrm{bg}}$ is estimated below. For non-relativistic axions with
$v_a\sim10^{-3}$, the emission is Doppler broadened to
$\Delta\nu/\nu_\gamma\sim10^{-3}$, implying $\Delta\nu\simeq100~\mathrm{MHz}$ at
$\nu_\gamma=100~\mathrm{GHz}$. Broadband observations would therefore dilute the
signal by a factor $\Delta\nu/\Delta\nu_{\rm inst}$, making high-resolution
spectroscopic observations essential. From an observational standpoint, the relevant figure of merit is the
signal-to-noise ratio per spectral channel \citep{2022PASP..134g4501A},
\begin{equation}
	\mathrm{SNR} \simeq \frac{F_{\nu,a\gamma}}{\sigma_\nu},
\end{equation}
where $\sigma_\nu$ denotes the rms noise in a channel of width $\Delta\nu$ for a
given integration time. A detectable feature requires both a sufficient
signal-to-noise ratio and a non-negligible spectral contrast $\delta$ with
respect to the local continuum.

To estimate the expected background continuum level, we follow
\citet{lyuti10.1093/mnras/sty3303} and parametrize the precursor electromagnetic
power emitted in radio bands by a fraction $\eta_R$ of the available
electromagnetic luminosity. For an equal-mass, magnetized BNS
the characteristic luminosity can be written as
\begin{equation}
	L_2(d) \sim
	\frac{B^2\, G m R^6}{c\, d^5},
\end{equation}
leading to a radio flux density
\begin{equation}
	F_R(d) \sim
	\eta_R\,
	\frac{L_2(d)}{4\pi D^2 \nu}
	=
	\eta_R\,
	\frac{B^2\, G m R^6}
	{4\pi c\, D^2\, \nu}\,
	d^{-5}.
\end{equation}

In the following, the background flux density is taken to be the radio precursor emission, $F_\nu^{\mathrm{bg}} \equiv F_R$. 

For representative neutron-star parameters
$m=1.4\,M_\odot$, $R=12~\mathrm{km}$,
$B=10^{13}~\mathrm{G}$, a source distance $D=10~\mathrm{kpc}$, observing
frequency $\nu=100~\mathrm{GHz}$, and $\eta_R=10^{-5}$, the resulting continuum
flux density spans
\begin{equation}
	F_\nu^{\rm bg} \sim 6\times10^{1}~\mathrm{Jy}
	\quad (d\simeq300~\mathrm{km}),
	\qquad
	F_\nu^{\rm bg} \sim 1.5\times10^{-1}~\mathrm{Jy}
	\quad (d\simeq1000~\mathrm{km}).
\end{equation}
For axion-induced line fluxes in the range
$F_{\nu,a\gamma}\sim1$--$10~\mu\mathrm{Jy}$, this corresponds to fractional
contrasts
\begin{equation}
	\delta \sim 2\times10^{-8}\text{--}2\times10^{-7}
	\quad (d\simeq300~\mathrm{km}),
	\qquad
	\delta \sim 7\times10^{-6}\text{--}7\times10^{-5}
	\quad (d\simeq1000~\mathrm{km}).
\end{equation}

By contrast, the signal-to-noise ratio per channel depends only on the
instrumental sensitivity. For rms noise levels
$\sigma_\nu\sim1$--$5~\mu\mathrm{Jy}$ in a
$\Delta\nu\simeq100~\mathrm{MHz}$ channel, one finds
$\mathrm{SNR}\sim0.2$--$10$ for $F_{\nu,a\gamma}=1$--$10~\mu\mathrm{Jy}$. These
estimates indicate that, while thermal-noise–limited detections may be feasible
at millimeter frequencies, the detectability of the axion-induced signal during
the late inspiral remains limited by its small contrast with the bright
precursor continuum. Remarkably, for the benchmark coupling $g_{a\gamma}=10^{-11}\,\mathrm{GeV^{-1}}$, the
axion-induced line flux increases to
$F_{\nu,a\gamma}\sim0.1$--$1~\mathrm{mJy}$ at $\nu=100~\mathrm{GHz}$. This
corresponds to fractional contrasts
$\delta\sim10^{-6}$--$10^{-5}$ at $d\simeq300~\mathrm{km}$, where the signal
remains effectively undetectable due to the bright precursor continuum, and
$\delta\sim10^{-4}$--$10^{-2}$ at $d\simeq1000~\mathrm{km}$, where the contrast
and signal-to-noise ratio are sufficient for detectability with targeted
millimeter observations. The expected merger rate at the distances considered here should be interpreted in
terms of the Galactic rate,
$\mathcal{R}_{\rm MW}\sim10^{-5}$--$10^{-4}\,\mathrm{yr^{-1}}$,
appropriate for a single Milky Way–like galaxy. Extrapolation to larger
volumes requires accounting for the local galaxy number density,
$n_{\rm gal}\sim10^{-2}\,\mathrm{Mpc^{-3}}$, which converts the per-galaxy rate
into a volumetric one and determines the number of mergers expected within a
given observational horizon.

\section{Conclusions}
\label{conclude}

We have presented an approximate framework for resonant axion--photon conversion in BNS magnetospheres during their early inspiral phase. Modeling the time-slice binary electromagnetic environment as the superposition of two rotating stellar dipolar fields, we show that the conversion is controlled by extended resonant surfaces whose dynamical geometry sets the emitted power. The inspiral-driven evolution of these surfaces imprints a distinctive spectral--temporal structure on the electromagnetic signal, intrinsically correlated with the gravitational wave emission.

Binary neutron star early inspirals thus emerge as time-dependent, multimessenger probes of axion physics. We find the hundred-GHz signal is narrowband and can be characterized by the spectral flux density \(F_\nu(\nu_\gamma,f_{\rm GW})\), which directly links the electromagnetic emission to the binary dynamics. The axion mass $m_a \in [50,170] \,\rm \mu eV$ combined with photon coupling $g_{a\gamma} \lesssim 10^{-11}\,\mathrm{GeV^{-1}}$ seems the suitable parameter region  to observe these effects. Gravitational wave frequencies detectable by DECIGO/BBO and ET/CE can be correlated for stellar separations $d\in[400,1000]$, km under the quasi-circular evolution in the early inspiral. This enables targeted electromagnetic searches, allowing direct comparison with the sensitivities of current and forthcoming radio and millimeter facilities such as ALMA, SKA and VLA. Future work should address inclined magnetic moments, misaligned spins, and more realistic plasma configurations, as well as relax the quasi-static inspiral approximation. These extensions will further assess the robustness and observability of axion-induced signals from compact binaries.

\begin{acknowledgments}
	We thank D. Barba-González and F. Giacchino for useful discussions. This work was supported by Junta de Castilla y León projects SA101P24, SA091P24, MICIU project PID2022-137887NB-I00, Gravitational Wave Network (REDONGRA) Strategic Network (RED2024-153735-E) from Agencia Estatal de Investigación del MICIU (MICIU/AEI/10.13039/501100011033).
\end{acknowledgments}

\bibliography{refBNSaxion}
\bibliographystyle{apsrev4-2}

\appendix
\section*{Details of the BNS magnetic field model}
\phantomsection
\label{ap:calculos}

In this appendix we collect the electromagnetic field expressions used
throughout our analysis.
We begin by recalling the electromagnetic field generated by a single rotating
neutron star in vacuum \citep{Deutsch1955}. Each star is modeled as a perfectly
conducting sphere of radius $a$, endowed with a magnetic dipole whose axis is
inclined by an angle $\chi$ with respect to the rotation axis. The stellar spin
angular frequency is denoted by $\omega$. In the near-zone regime,
$r \ll R_{\rm LC}\equiv c/\omega$, retardation effects are subdominant and the
fields admit simple analytical expressions.

In spherical coordinates $(r,\theta,\varphi)$, the magnetic field components of
an oblique rotator take the form
\begin{align}
B_r &= R_1(a)\,\frac{a^3}{r^3}
\left(\cos\chi\cos\theta+\sin\chi\sin\theta\cos\lambda\right),\\
B_\theta &= \frac{1}{2}R_1(a)\,\frac{a^3}{r^3}
\left(\cos\chi\sin\theta-\sin\chi\cos\theta\cos\lambda\right),\\
B_\varphi &= \frac{1}{2}R_1(a)\,\frac{a^3}{r^3}\,
\sin\chi\,\sin\lambda,
\end{align}
where $\lambda\equiv\varphi-\omega t$ is the rotational phase of the dipole.
The factor $R_1(a)$ fixes the magnetic-field normalization at the stellar
surface; in the aligned limit ($\chi=0$) one has $R_1(a)\equiv B_{\rm p}$, the
polar surface magnetic field.

In the same near-zone approximation, the corresponding electric field is
\begin{align}
E_r &= -\frac{1}{4}\,\frac{\omega a}{c}\,R_1(a)\,\frac{a^4}{r^4}
\left[\cos\chi\,(3\cos 2\theta+1)+3\sin\chi\,\sin 2\theta\,\cos\lambda\right],\\
E_\theta &= -\frac{1}{2}\,\frac{\omega a}{c}\,R_1(a)\,\frac{a^2}{r^2}
\left[\frac{a^2}{r^2}\cos\chi\,\sin 2\theta
+\sin\chi\left(1-\frac{a^2}{r^2}\cos 2\theta\right)\cos\lambda\right],\\
E_\varphi &= \frac{1}{2}\,\frac{\omega a}{c}\,R_1(a)\,\frac{a^2}{r^2}
\left(1-\frac{a^2}{r^2}\right)\sin\chi\,\cos\theta\,\sin\lambda.
\end{align}
These expressions capture the leading-order electromagnetic fields of an oblique
rotator in the near-zone.

We now consider a binary system composed of two neutron stars of this type,
separated by a distance $d$. The total electromagnetic field is obtained by
superposing the individual stellar contributions.

For the companion star, we introduce a translated coordinate system
$(r',\theta')$ in the meridional plane, with the origin displaced by a distance
$d$ along the equatorial direction. The relation between the two coordinate
systems is
\begin{align}
r' &= \sqrt{r^2+d^2+2dr\sin\theta},\\
\theta' &= \operatorname{atan2}(r\sin\theta+d,\; r\cos\theta),
\end{align}
where $\operatorname{atan2}(y,x)$ denotes the two-argument arctangent, ensuring
the correct angular quadrant.

With these definitions, the relevant trigonometric functions read
\begin{equation}
\cos\theta'=\frac{r\cos\theta}{r'},\qquad
\sin\theta'=\frac{r\sin\theta+d}{r'}.
\end{equation}

The magnetic and electric fields of the companion star are obtained by evaluating
the single-star expressions at $(r',\theta')$ with the corresponding stellar
parameters $(a^\star,\omega^\star,\chi^\star)$. 

the magnetic field generated by the companion star in the spherical coordinate system centered on star $a^\star$. Throughout this section, the coordinate system is centered on star $a^\star$, while $a$ denotes the companion neutron star whose magnetic field contribution is re-expressed at a distance $d$ from the origin.

Restricting to the meridional plane, we consider a rigid translation of the
coordinate origin by a distance $d$ along the $x$--direction, where in the
original spherical coordinates $x = r \sin\theta, z = r \cos\theta$.  The coordinates $(r',\theta')$ associated with the translated origin are defined by $x' = d + r \sin\theta, z' = r \cos\theta$ so that
\begin{equation}
r' = \sqrt{x'^2 + z'^2}
   = \sqrt{d^2 + r^2 + 2 r d \sin\theta}.
\end{equation}

The polar angle $\theta'$ is computed using the two--argument arctangent,
\begin{equation}
\theta' = \operatorname{atan2}(x',z')
        = \operatorname{atan2}\!\big(d + r \sin\theta,\; r \cos\theta\big),
\end{equation}
which ensures the correct quadrant of $\theta'$ across the entire meridional
plane and avoids spurious discontinuities that would arise from using the
single--argument $\arctan$.

With this choice of coordinates, the non--vanishing components of the
transformed vector field read
\begin{align}
    B_r^{\star} =& \Lambda_B(r,\theta)
    \Bigg[
        2 r^2 \cos^3\theta \cos\chi^{\star} (d + r \sin\theta)
        + \cos\theta \cos\chi^{\star} (d + r \sin\theta)
        \big(-d^2 + r^2 - r^2 \cos 2\theta + r d \sin\theta \big) \nonumber \\
    &\quad
        + \frac{1}{2} (d + r \sin\theta)
        \big( -r d (-7 + \cos 2\theta)
        + 4 (d^2 + r^2) \sin\theta \big) \sin\chi^{\star}
    \Bigg], \\
    B_\theta^{\star} =& \Lambda_B(r,\theta)
    \Bigg[
        \cos\chi^{\star} \sin\theta (d + r \sin\theta)^3
        + r \cos^2\theta \cos\chi^{\star} (d + r \sin\theta)(3d + r \sin\theta)
        - d r^2 \cos^3\theta \sin\chi^{\star} \nonumber \\
    &\quad
        + \cos\theta \big( 2d^3 - r (-3d^2 + r^2) \sin\theta \big) \sin\chi^{\star}
    \Bigg], \\
    B_\phi^{\star} =& 0 ,
\end{align}
where
\begin{equation}
\Lambda_B(r,\theta)
= \frac{a^{\star 3} R_1(a^{\star})}
       {2 (d + r \sin\theta)\, (d^2 + r^2 + 2 r d \sin\theta)^{5/2}} .
\end{equation}

The apparent singularity at $d + r \sin\theta = 0$ corresponds to the displaced
axis $x' = 0$ and is a coordinate artifact of the chosen meridional
parametrization.

The total electromagnetic field
of the binary system is then
\begin{equation}
\mathbf{B}_{\rm binary}=\mathbf{B}+\mathbf{B}^\star,\qquad
\mathbf{E}_{\rm binary}=\mathbf{E}+\mathbf{E}^\star,
\end{equation}
which describes the electromagnetic configuration of the binary system in the
meridional plane at separation $d$ within the vacuum dipole model.

\end{document}